\documentclass[number]{elsarticle}
\journal{Physics Letters B}

\usepackage[english]{babel}
\usepackage{amsmath,amssymb,amsfonts,amstext}
\usepackage{url}
\usepackage{color,lscape}
\usepackage{float}

\usepackage{bm}
\renewcommand{\v}[1]{\bm{ #1 }}

\begin{document}  
\begin{frontmatter}

\title{Bayesian model selection for electromagnetic kaon production on the nucleon}
\author[gent]{L.~De~Cruz\corref{author}}\ead{Lesley.DeCruz@UGent.be}
\author[glasgow]{D.~G.~Ireland}
\author[gent]{P.~Vancraeyveld}
\author[gent]{J.~Ryckebusch}
\address[gent]{Department of Physics and Astronomy, Ghent University, Proeftuinstraat 86, B-9000 Gent, Belgium}
\address[glasgow]{Department of Physics and Astronomy, University of Glasgow, Glasgow G12 8QQ, United Kingdom}
\cortext[author]{Corresponding author}

\begin{abstract}	
We present the results of a Bayesian analysis of a Regge model to describe the background contribution for $K^+\Lambda$ and $K^+\Sigma^0$ photoproduction. The model is based on the exchange of $K^{+}(494)$ and $K^{\ast+}(892)$ trajectories in the $t$-channel. We utilise the Bayesian evidence $\mathcal{Z}$ to determine the best model variant for each channel. The Bayesian evidence integrals were calculated using the Nested Sampling algorithm. For different prior widths, we find decisive Bayesian evidence ($\Delta \ln \mathcal{Z} \approx 24$) for a $K^+\Lambda$ photoproduction Regge model with a positive vector coupling and a negative tensor coupling constant for the $K^{\ast+}(892)$ trajectory, and a rotating phase factor for both trajectories. Using the $\chi^2$ minimisation method, one could not draw this conclusion from the same dataset. For the $K^+\Sigma^0$  photoproduction Regge model, on the other hand, the difference between the evidence integrals is insufficient to pinpoint one model variant. 
\end{abstract}

\end{frontmatter}

\section{Introduction}

To resolve the structure of the nucleon, many models based on effective degrees of freedom have been developed. One of the more popular ones is the constituent quark model (CQM), which presents the nucleon as a system of three constituent quarks~\cite{capstick-2000}. The CQM, however, predicts far more resonances than confirmed by experiment. This may lead one to turn to other models that predict fewer resonances~\cite{anselmino-1992}.

Alternatively, this discrepancy may arise because missing resonances do not couple to the channels commonly used for nucleon spectroscopy, such as the pion-nucleon ($\pi N$) channel. This issue can be addressed by using an electromagnetic probe instead of a pion, and by examining decay channels other than $\pi N$. Hence, electromagnetic open strangeness or kaon-hyperon ($KY$) production is suggested as a key process to seek for unobserved resonances. The analysis of this process faces other difficulties, such as a small cross section and a high threshold~\cite{saghai-2006, corthals-2006}. Unlike threshold one-pion photoproduction, which is dominated by the $\Delta(1232)$ resonance, the $KY$ reaction channel opens in a resonance-rich energy region. The identification of those resonances constitutes a major challenge in modelling $KY$ production. A second characteristic of this channel is the great importance of the non-resonant or background contributions. Hence, a correct determination of the background is crucial for a correct assessment of the resonance contributions.

The earliest studies of the $KY$ production focused on the estimation of coupling constants within a single model variant~\cite{adelseck-1986, adelseck-1988}. Only when the emphasis came to lie on identifying missing resonances did the focus shift from parameter estimation to model comparison. The statistical tools, however, have not been adapted to this new objective. The least-squares method in particular has often been stretched beyond its limits, being used not only as an optimisation tool, but also as a model selection criterion. In this Letter, we present the Bayesian evidence computation, a method based on the principles of Bayesian inference, as a more robust and well-founded tool for model comparison. 

The outline of this Letter is as follows. The next section introduces the Bayesian evidence and the Nested Sampling (NS) algorithm for evidence computation. The effectiveness of this algorithm is subsequently demonstrated for a Regge model in Section \ref{sec:Regge} and the results are discussed in Section \ref{sec:Results}.
The conclusions and outlook are given in Section \ref{sec:conclusion}.

\section{Bayesian analysis}\label{sec:Bayes}

Bayesian analysis is an established tool for model selection in astronomy and cosmology, and is gaining momentum in other fields~\cite{mukherjee-2006,feroz-2008}. The potential of this method in hadronic physics was recently demonstrated in a Bayesian analysis of pentaquark data by Ireland et al.~\cite{ireland-2008}, and for parameter estimation in effective field theories by Schindler et al.~\cite{schindler-2008}. The quantity of interest for model comparison is the Bayesian evidence, which will be derived below. 

One can straightforwardly derive the posterior probability of a model $M$, given a set of experimental data $\left\lbrace d_k \right\rbrace$. Indeed, using Bayes' theorem, one can write this probability as
\begin{equation}
P(M|\left\lbrace d_k \right\rbrace) = \frac{ P(\left\lbrace d_k \right\rbrace|M) \, P(M)}{P(\left\lbrace d_k \right\rbrace)}.
\end{equation}
The quantity $P(\left\lbrace d_k \right\rbrace|M)$ is referred to as the marginal likelihood or the Bayesian evidence ($\mathcal{Z}$). If the model $M$ can be parametrised with a set $\bm{\alpha_M}$, this probability can be written as an integral over all possible values of these parameters. This procedure, which is referred to as marginalisation, yields the following expression for the Bayesian evidence:
\begin{align}
\mathcal{Z} &\equiv P(\left\lbrace d_k \right\rbrace|M)\label{eqn:evidence1} \\
		&= \int P(\left\lbrace d_k \right\rbrace,\v{\alpha_M}|M)\, d\v{\alpha_M} \label{eqn:evidence2}\\
		&= \int \underbrace{P(\left\lbrace d_k \right\rbrace|\v{\alpha_M},M)}_{(i)} \, \underbrace{P(\v{\alpha_M}|M)}_{(ii)}\, d\v{\alpha_M}. \label{eqn:evidence}
\end{align}
Eq.~(\ref{eqn:evidence}) states that the Bayesian evidence is the integral of the product of two distributions: $(i)$ the probability of the dataset $\left\lbrace d_k \right\rbrace$, given the set of parameters $\v{\alpha_M}$ and the model $M$, and $(ii)$ the probability of the set of parameters $\v{\alpha_M}$, given the model $M$. The first factor, $P(\left\lbrace d_k \right\rbrace|\v{\alpha_M},M)$, can be identified as the likelihood function, $\mathcal{L}(\v{\alpha_M})$. Any prior knowledge of the parameters' probability distribution before considering the data $\left\lbrace d_k\right\rbrace$ is contained in the second factor $P(\v{\alpha_M}|M)$. This distribution, which is indispensable in Bayesian statistics, is referred to as the prior distribution $\pi(\v{\alpha_M})$. These two substitutions allow us to write the evidence in a more familiar form,
\begin{align}
\mathcal{Z} &= \int \mathcal{L}(\v{\alpha_M})\, \pi(\v{\alpha_M})\, d\v{\alpha_M} \label{eqn:full-evidence}, 
\end{align}
in which the explicit dependence on $\left\lbrace d_k \right\rbrace$ and $M$ is omitted for brevity.

It is clear that the actual quantity of interest for model comparison is the relative probability of a model $M_A$ versus a model $M_B$, given the available experimental data $\left\lbrace d_k \right\rbrace$. 
Writing down the probability ratios and subsequently applying Bayes' theorem, one can see how the evidence emerges from this expression:

\begin{align}
 \frac{P(M_A|\left\lbrace d_k \right\rbrace)}{P(M_B|\left\lbrace d_k \right\rbrace)} &= \frac{P(\left\lbrace d_k \right\rbrace|M_A)}{P(\left\lbrace d_k \right\rbrace|M_B)} \,  \frac{P(M_A)}{P(M_B)}\label{eqn:explain_evidence}\\
	&= \frac{\mathcal{Z_A}}{\mathcal{Z_B}} \;\; \mathrm{for}\, P(M_A) = P(M_B).
\end{align}

	\begin{table}
		\small
		\begin{center}
		\caption{Jeffreys' scale for the natural logarithms of evidence ratios $\Delta \ln\mathcal{Z} = \ln\frac{\mathcal{Z_A}}{\mathcal{Z_B}}$~\cite{jeffreys-1961, kass-1995}.}
		\label{tab:jeffreys}
			\begin{tabular}{rcll}
				\hline
				&$|\Delta \ln\mathcal{Z}|$&$ < 1$ 		& Not worth more than a bare mention\\
				$ 1 < $&$|\Delta \ln\mathcal{Z}|$&$ < 2.5 $	& Significant \\
				$ 2.5 < $&$|\Delta \ln\mathcal{Z}|$&$ < 5$ 	& Strong to very strong\\
				$ 5 < $&$|\Delta \ln\mathcal{Z}|$&    		& Decisive\\
				\hline
			\end{tabular}
		\end{center}
	\end{table}

Any prior preference for one model over the other can be incorporated by the factor $\frac{P(M_A)}{P(M_B)}$. As we have no prior preference for one of the models, we can take this value to be one, hence reducing the comparison of two models to the calculation of the evidence ratio, which is often referred to as the Bayes factor. The direct relation between $\mathcal{Z}$ and a model's probability elucidates the term ``evidence'': if a model has a higher value of $\mathcal{Z}$, there is more evidence in favour of this model. In accordance with our intuitive notions, evidence is not only based on experimental data, but also on theoretical restrictions that are incorporated through the prior distribution. The natural logarithm of the evidence ratio can be interpreted qualitatively with the aid of Jeffreys' scale, listed in Table \ref{tab:jeffreys}.

The analytical form of the likelihood function $\mathcal{L}(\v{\alpha_M})$ is rarely known and a normal distribution is often used to approximate it. 
Indeed, data points are independent and are usually reported to have normally distributed errors. This gives rise to a $\chi^2$-distribution for the quantity defined as
\begin{equation}
\chi^2(\v{\alpha_M}) = \sum_i^N \dfrac{\left( d_i-f_i(\v{\alpha}_M)\right)^2}{2 \sigma_i^2},
\end{equation}
where $\sigma_i$ is the error bar of data point $d_i$, and $f_i(\v{\alpha}_M)$ is the corresponding model prediction.
The $\chi^2$-distribution can be approximated by a normal distribution if the number of degrees of freedom $k$ is sufficiently large. This value is defined as $k = N - \dim(\v{\alpha}_M)$, which approximates the number of data points for a sufficiently large dataset and low number of free parameters. The expression for this limiting distribution is~\cite{barlow-1989}:
\begin{equation}
 \mathcal{L}(\v{\alpha_M}) \approx \frac{1}{2\sqrt{\pi k}}\exp{-\frac{(\chi^2(\v{\alpha_M}) - k)^2}{4k}}.
\end{equation}

\begin{figure}[]
		\begin{center}
			\includegraphics[height=.3\textheight]{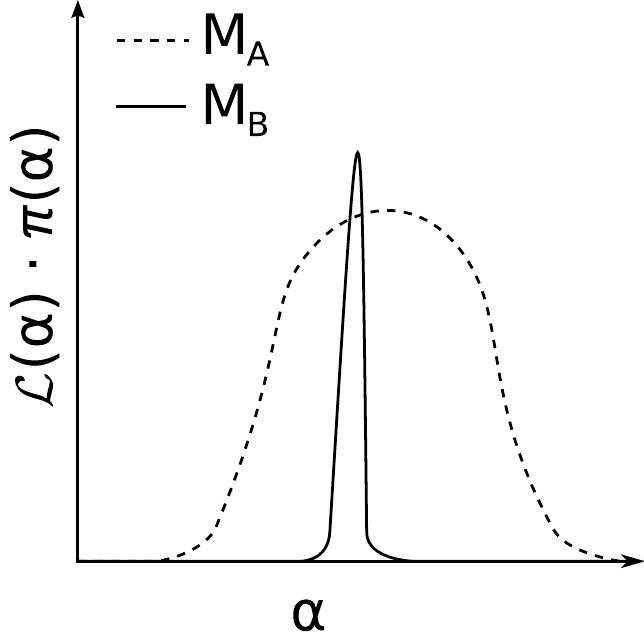}
			\caption{$\mathcal{L}(\alpha) \cdot \pi(\alpha)$ for two different single-parameter models $M_A$ and $M_B$. The traditional least-squares method would favour $M_B$, as it only takes into account the maximum value of the likelihood. In contrast, a Bayesian approach would favour model $M_A$, as its evidence, given by Eq.~(\ref{eqn:full-evidence}), is greater.
			\label{fig:bayes_peak}}
		\end{center}
	\end{figure}

In an analysis based on  $\chi^2$ minimisation -- which is an approximation to a maximum likelihood fit --  only the maximum value of $\mathcal{L}(\v{\alpha_M}) $ is considered for model selection. A Bayesian approach is more comprehensive, 
as it evaluates the model over its entire parameter space, and takes the prior distribution into account.  This distinction is illustrated in Fig. \ref{fig:bayes_peak}. 

Determining the evidence of a model is not a straightforward task, because it requires the calculation of multidimensional integrals of the type (\ref{eqn:full-evidence}). Most often, analytical simplifications are not possible and it is key to adopt numerical integration techniques that are optimised for the problem at hand. 

Nested Sampling (NS) is a novel integration technique for computing Bayesian evidence, developed by Skilling~\cite{sivia-2006, skilling-2006}. This technique significantly reduces the computational cost of the integral over the model's parameter space by transforming it into a one-dimensional integral over the prior mass $dX = \pi(\v{\alpha_M}) d\v{\alpha_M}$. This is accomplished by regarding the prior mass as a monotonically decreasing function of the likelihood, $\lambda$:
\begin{equation}
 X(\lambda) = \int_{\mathcal{L}(\v{\alpha_M}) > \lambda} \pi(\v{\alpha_M}) d\v{\alpha_M}.
\end{equation}

\begin{figure}[]
	\begin{center}
		\includegraphics[height=.35\textheight]{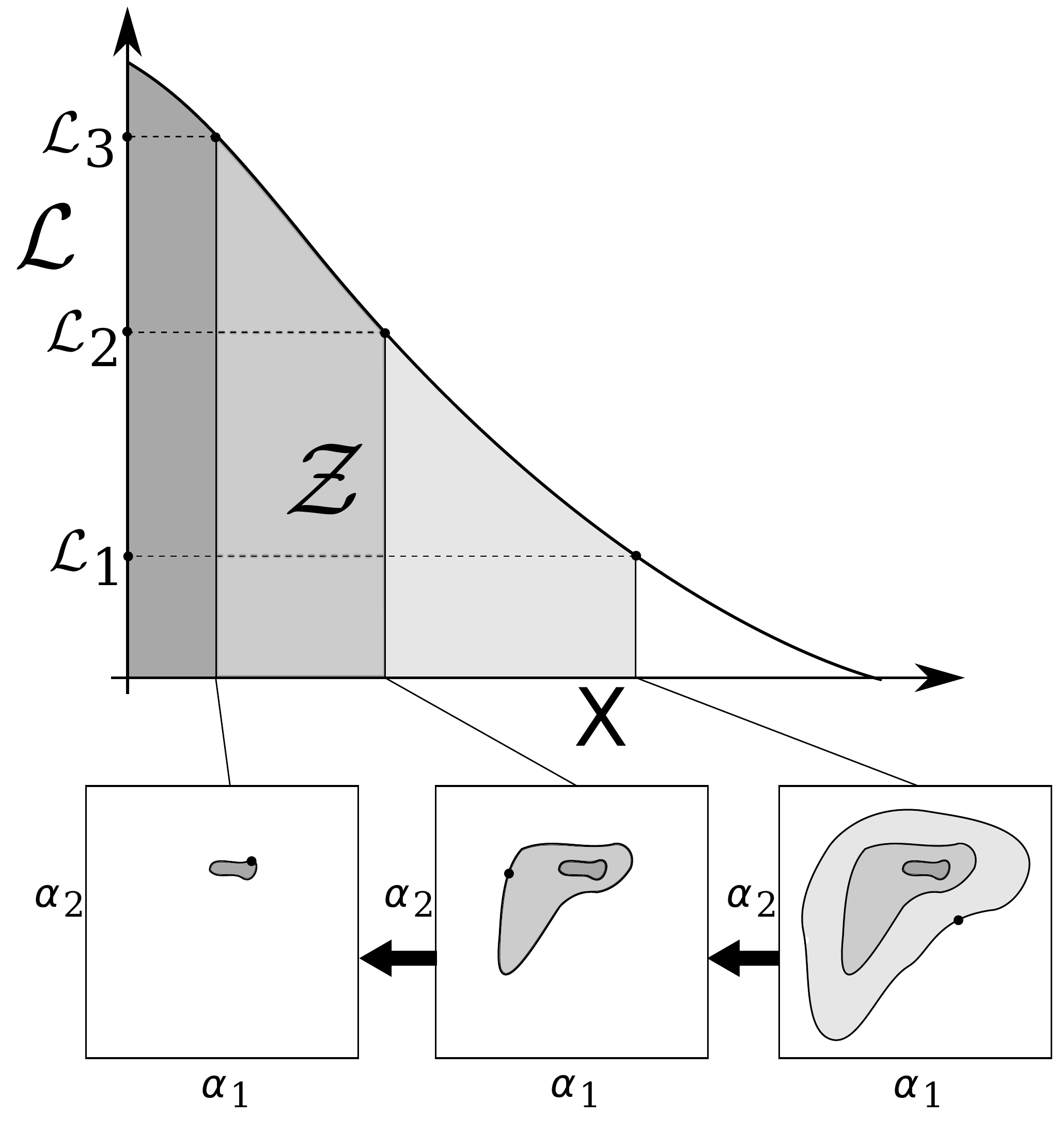}
		\caption{The Bayesian evidence $\mathcal{Z}$ as an integral of the likelihood $\mathcal{L}$ over the prior mass $X$. The lower plots illustrate that for a uniform prior, the increasing prior masses $X(\mathcal{L}_i)$ are equal to the area or mass $ \left\lbrace \, \v{\alpha} \, |\,  \mathcal{L(\v{\alpha})} > \mathcal{L}_i \right\rbrace $ inside nested iso-likelihood contours. The thick arrows indicate the order of integration. }
		
		\label{fig:ns}
	\end{center}
\end{figure}

Assuming a normalised prior, we can hence write the evidence as the following integral over $\mathcal{L}(X)$, the inverse of $X(\lambda)$,
\begin{equation}
 \mathcal{Z} = \int_0^1 \mathcal{L}\left(X\right) dX,
\end{equation}
as illustrated in Fig. \ref{fig:ns}.
This integral can be approximated by a sum of likelihoods, weighted with their respective prior mass contribution $\Delta X$. In NS, this sum is computed using Markov chain Monte Carlo methods to sample from the parameter space with the constraint $\mathcal{L}_{i+1} > \mathcal{L}_i$. The prior mass of a sample of $N$ points with likelihood greater than $\mathcal{L}_i$ can be estimated by a factor $\exp{-\frac{i}{N}}$. This can be derived from the probability distribution of the largest value $t_{N}$ of a set of $N$ uniformly distributed points in the interval $[0,1]$ (representing the total prior weight):
\begin{equation}
P_N(t) = N t^{N-1}.
\end{equation}
Indeed, the estimation value for $\left<\log{t}\right>_N$  is $ -\frac{1}{N}$, which after $i$ iterations amounts to a sum of $-\frac{i}{N}$.

Besides efficiently calculating the evidence, NS can also be used for determining the posterior distribution and, more importantly, for parameter estimation. 
A more detailed account of the technique as well as implementation examples can be found in Skilling's work~\cite{skilling-2006}.

In the following section, we will present the results of an application of the NS technique to a Regge model for $K^+\Lambda$ and $K^+\Sigma^0$ production.

\section{Bayesian analysis of a Regge Model} \label{sec:Regge}

Regge phenomenology is a powerful tool to economically describe reactions at high energies~\cite{collins-1977}. A Regge model based on the exchange of charged meson Regge trajectories was applied to $KY$ photoproduction by Vanderhaeghen, Guidal and Laget~\cite{vanderhaeghen-1997}. They found the exchange of two $t$-channel trajectories sufficient to successfully describe the cross sections as well as polarisation observables in photoproduction of kaons above the resonance region~\cite{guidal-2003}.

In the Regge-plus-resonance (RPR) description of electromagnetic $KY$ production, developed by Corthals et al., the Regge background is complemented with $s$-channel nucleon ($N^{\ast}$) resonances. This hybrid approach ensures a correct high-energy behaviour as well as an improved description of the resonance region~\cite{corthals-2006, corthals-2007a, corthals-2007b}.

In the case of $K^+\Lambda$ and $K^+\Sigma^0$ photoproduction, the Regge background can be modelled with the exchange of the $K^{+}(494)$ and $K^{\ast+}(892)$ trajectories. This amplitude is derived from the $t$-channel Feynman amplitude by replacing the Feynman propagator by the respective Regge propagator~\cite{corthals-2006}:
\begin{equation}
\begin{split} 
 \mathcal{P}^{K^+}_{Regge}(s,t) =& \left(\frac{s}{ s_0}\right)^{\alpha_{K}(t)} \frac{1}{\sin\bigl(\pi\alpha_{K}(t)\bigr)} \\
 \times& \frac{\pi \alpha'_K}{\Gamma\bigl(1+\alpha_{K}(t)\bigr)} \ \left\{ \begin{array}{c}
1 \\ e^{-i\pi\alpha_{{K}}(t)} 
\end{array}\right\} \,
\label{eq:reggeprop_K}
\end{split}
\end{equation}

\begin{equation}
\begin{split} 
 \mathcal{P}^{K^{\ast+}}_{Regge}(s,t) =& \left(\frac{s}{ s_0}\right)^{\alpha_{K^{\ast}}(t)-1} \frac{1}{\sin\bigl(\pi\alpha_{K^{\ast}}(t)\bigr)}  \\
 \times&  \frac{\pi \alpha'_{K^{\ast}}}{\Gamma\bigl(\alpha_{K^{\ast}}(t)\bigr)} \ \left\{ \begin{array}{c}
1 \\ e^{-i\pi\alpha_{{K^{\ast}}}(t)} 
\end{array}\right\}. \,
\label{eq:reggeprop_Ks}
\end{split}
\end{equation}
The kaon trajectories are given by~\cite{corthals-2006}:
\begin{align}
\alpha_K (t) &=  0.70 \; \mathrm{ GeV}^{-2} \left( t - m_K^2\right)\\
\alpha_K^{*} (t) &=  1 + 0.85 \; \mathrm{ GeV}^{-2} \left( t - m_{K^*}^2\right),
\end{align}
and the scale factor $s_0 = 1 \;\mathrm{ GeV}^{-2}$.
The so-called sign factor in the Regge propagator is reduced to a phase factor of either 1 (constant phase) or $e^{-i\pi\alpha(t)}$ (rotating phase) due to the strong degeneracy of the trajectories. This assumption is inspired by the structureless high-energy differential cross-section.
These phases cannot be determined on theoretical grounds. The possibility of the $K^{+}$ and $K^{\ast+}$ trajectories having a constant phase is excluded, as this combination gives rise to a photon asymmetry $\Sigma = 0$, which disagrees with the data. The remaining three possibilities, namely rotating $K^{+}$ /rotating $K^{\ast+}$, rotating $K^{+}$/constant $K^{\ast+}$, and constant $K^{+}$/rotating $K^{\ast+}$, will be abbreviated as rot./rot., rot./cst. and cst./rot., respectively.

Apart from the three choices with regard to the phases, the model has three continuous parameters. These are the strong coupling constant $g_{K^+Y p}$ of the $K^{+}$ 
trajectory and the tensor and vector couplings of the $K^{\ast +}$ trajectory,
\begin{equation}
G_{K^{\ast +}}^{v,t} = \frac{e \, g_{{K^{\ast +}}\, Y p}^{v,t}}{ 4 \pi} \ \kappa_{K^+{K^{\ast +}}}\;.\label{eq: bg_free_pars} 
\end{equation}
Here, $ \kappa _ { K^{+} K ^{ *+}} $ is the transition magnetic moment for $ K ^{*+} \rightarrow \gamma K ^{+}$ decay.

In the following analysis, the Regge background will be constrained by the high-energy data, in accordance with the RPR-approach. In this energy region, there is a set of 72 data points for the $K^+\Lambda$ channel, comprising 56 differential cross section data points ($\frac{d\sigma}{dt}$)~\cite{boyarski-1969}, 9 photon asymmetries ($\Sigma$)~\cite{quinn-1979} and 7 recoil asymmetries ($P$)~\cite{vogel-1972}. The database for $K^+\Sigma^0$ photoproduction at high energies is even smaller, with only 48 differential cross section data points~\cite{boyarski-1969} and 9 photon asymmetries ($\Sigma$)~\cite{quinn-1979}. Optimisation of the above-mentioned parameters against these data reveals that there are several model variants with comparable $\chi^2$ values~\cite{corthals-2006}. For example, in both $K^+\Lambda$ and $K^+\Sigma^0$ photoproduction, the signs of $G^{v}_{K^{\ast+}}$ and $G^{t}_{K^{\ast+}}$ cannot be established conclusively using the $\chi^2$-method~\cite{corthals-2006}.

The sign and phase ambiguities may not seem important for the Regge model itself. For the RPR model, however, an exact determination of the background parameters is of major importance, because it affects the extraction of the resonance information. 

Ref.~\cite{corthals-2007b} shows that by comparing the RPR model variants for electromagnetic $K^{+}\Lambda$ production to photo- and electroproduction data from the resonance region, all but one Regge background model can be eliminated. We will show that the Bayesian evidence can be used to distinguish among the twelve different models that result from the possible sign and phase combinations, using only the high-energy dataset.

\section{Results}\label{sec:Results}

The prior $\pi(\v{\alpha})$ is chosen to be a uniform distribution. Note that under conditions of highly concentrated likelihood, for which the prior distribution varies mildly, the likelihood dominates the shape of the posterior distribution~\cite{sivia-2006}. Accordingly, the evidence calculations will not be largely affected by the choice with regard to the prior distribution. This means that a uniform distribution will lead to results that are similar to those obtained with a Gaussian or any other well-behaved distribution.  We show that the bulk of the likelihood is indeed concentrated at parameter values below 100 by demonstrating that evidence calculations for prior widths equal to 100 and much greater than 100 yield the same results.

Prior information exists for the coupling constants of the $K^+ Y^{0} p$ vertices. Indeed, the following relations follow from  SU(3) symmetry~\cite{adelseck-1990}:
\begin{align}
	g_{K^+\Lambda p} 	& = -\frac{1}{\sqrt{3}}(3-2\alpha)\,g_{\pi NN} \\
	g_{K^+\Sigma^{0} p}	& = (2\alpha-1)\,g_{\pi NN}
\end{align}
where $\alpha = F/(F+D)$ quantifies the ratio of F-type to D-type coupling and $g_{\pi NN}$  is the pion-nucleon coupling constant. It is commonly assumed that SU(3) symmetry can be broken at the 20\% level~\cite{donoghue-1982}. Inserting the experimentally determined values $\alpha  = 0.644$  and $g^2_{\pi NN}/\sqrt{4\pi} = 14.3$ yields the following prior ranges for the coupling constants of the $K^+ Y^{0} p$ vertices~\cite{donoghue-1982,adelseck-1990}:
\begin{align}
 -4.5 \leq &\frac{g_{K^+\Lambda p}}{\sqrt{4\pi}} \leq -3.0\\
 0.9 \leq &\frac{g_{K^+\Sigma^{0} p}}{\sqrt{4\pi}} \leq 1.3.
\end{align}

There are no reliable theoretical constraints for the $K^{\ast+}Y^{0} p$ vertices~\cite{guidal-phd}. We therefore choose a uniform distribution between zero and a value much larger than the natural value of one. To justify this choice for the prior interval, a sensitivity analysis of the evidence ratios is performed by repeating the calculations for different prior ranges.

The results of these calculations for the $K^+\Lambda$ and $K^+\Sigma^{0}$ production models are displayed in Table \ref{tab:lnz-kl} and \ref{tab:lnz-ks} respectively. These tables list the computed values of $\Delta\ln\mathcal{Z} = \ln\mathcal{Z} - \ln\mathcal{Z}_{max}$, using a prior width of respectively 100, 1000 and 10000. Changing the prior width from 100 to 1000 results in a difference of less than $5\%$ in the computed values. For a prior width of 10000, the error increases significantly due to a reduced sampling efficiency. More importantly, however, the ranking of the models is not significantly affected. Clearly, the effect of the prior width on the relative probabilities of the models is negligible, provided that it is large enough to contain the area where most of the likelihood is concentrated. 

\begin{table*} [htp] 
	\scriptsize	\centering
	\caption{Logarithms of the evidence ratios ($\Delta \ln{\mathcal{Z}} \equiv \ln{\left(\mathcal{Z}/\mathcal{Z}_{max}\right)}$)	for the twelve model variants resulting from phase and sign ambiguities in the two-trajectory Regge model for $K^{+}\Lambda$ photoproduction. The results are listed in order of decreasing probability for a prior width of 100.}
	\begin{tabular}{|ccc|rcr|rcr|rcr|}
	\hline
	$G^{v}$ & $G^{t} $& $K^{+} / K^{*+}$ phase & \multicolumn{3}{|c|}{ $\pi = U(0,\pm100)$} & \multicolumn{3}{|c|}{ $\pi = U(0,\pm1000)$} & \multicolumn{3}{|c|}{ $\pi = U(0,\pm10000)$}\\
	\hline
	\label{tab:lnz-kl}
	 $+$ &   $-$ &  rot. /  rot.  & 0\phantom{.00}  & & & 0\phantom{.00}  & & & 0 & & \\
	 $-$ &   $-$ &  rot. /  cst.  & $-24.30$ & $\pm$ & $0.75$ & $-24.7\phantom{0}$ & $\pm$ & $3.5\phantom{0}$ & $-24$ & $\pm$ & $41$ \\
	 $+$ &   $+$ &  rot. /  rot.  & $-77.23$ & $\pm$ & $0.76$ & $-77.3\phantom{0}$ & $\pm$ & $2.9\phantom{0}$ & $-79$ & $\pm$ & $59$ \\
	 $-$ &   $+$ &  rot. /  cst.  & $-387.04$ & $\pm$ & $0.77$ & $-387.7\phantom{0}$ & $\pm$ & $4.2\phantom{0}$ & $-411$ & $\pm$ & $68$ \\
	 $+$ &   $+$ &  rot. /  cst.  & $-2366.2\phantom{0}$ & $\pm$ & $1.3\phantom{0}$ & $-2374\phantom{.00}$ & $\pm$ & $17\phantom{.00}$ & $-2530$ & $\pm$ & $280$ \\
	 $-$ &   $-$ &  cst. /  rot.  & $-2870.66$ & $\pm$ & $0.73$ & $-2871.0\phantom{0}$ & $\pm$ & $4.0\phantom{0}$ & $-2890$ & $\pm$ & $81$ \\
	 $-$ &   $+$ &  cst. /  rot.  & $-3384.83$ & $\pm$ & $0.83$ & $-3386.7\phantom{0}$ & $\pm$ & $7.6\phantom{0}$ & $-3440$ & $\pm$ & $120$ \\
	 $+$ &   $-$ &  cst. /  rot.  & $-3475.74$ & $\pm$ & $0.73$ & $-3478.5\phantom{0}$ & $\pm$ & $5.4\phantom{0}$ & $-3526$ & $\pm$ & $81$ \\
	 $-$ &   $-$ &  rot. /  rot.  & $-4297.9\phantom{0}$ & $\pm$ & $1.6\phantom{0}$ & $-4317\phantom{.00}$ & $\pm$ & $32\phantom{.00}$ & $-4460$ & $\pm$ & $200$ \\
	 $-$ &   $+$ &  rot. /  rot.  & $-4952.1\phantom{0}$ & $\pm$ & $1.7\phantom{0}$ & $-4970\phantom{.00}$ & $\pm$ & $27\phantom{.00}$ & $-5180$ & $\pm$ & $230$ \\
	 $+$ &   $+$ &  cst. /  rot.  & $-5092.18$ & $\pm$ & $0.72$ & $-5094.2\phantom{0}$ & $\pm$ & $7.8\phantom{0}$ & $-5150$ & $\pm$ & $130$ \\
	 $+$ &   $-$ &  rot. /  cst.  & $-19602.3\phantom{0}$ & $\pm$ & $4.2\phantom{0}$ & $-19710\phantom{.00}$ & $\pm$ & $200\phantom{.00}$ & $-20430$ & $\pm$ & $440$ \\
	\hline
	\end{tabular} 
\end{table*}

\begin{table*} [htp] 
	\scriptsize	\centering
	\caption{Logarithms of the evidence ratios ($\Delta \ln{\mathcal{Z}} \equiv \ln{\left(\mathcal{Z}/\mathcal{Z}_{max}\right)}$)	for the twelve model variants resulting from phase and sign ambiguities in the two-trajectory Regge model for $K^{+}\Sigma^{0}$ photoproduction. The results are listed in order of decreasing probability for a prior width of 100.}
	\begin{tabular}{|ccc|rcr|rcr|rcr|}
	\hline
	$G^{v}$ & $G^{t} $& $K^{+} / K^{*+}$ phase & \multicolumn{3}{|c|}{ $\pi = U(0,\pm100)$} & \multicolumn{3}{|c|}{ $\pi = U(0,\pm1000)$} & \multicolumn{3}{|c|}{ $\pi = U(0,\pm10000)$}\\
	\hline
	\label{tab:lnz-ks}
	 $-$ &   $-$ &  rot. /  cst.  & 0\phantom{.00}  & & & 0\phantom{.00}  & & & 0\phantom{.00}  & & \\
	 $+$ &   $-$ &  rot. /  cst.  & $-0.50$ & $\pm$ & $0.42$ & $-0.42$ & $\pm$ & $0.63$ & $-0.55$ & $\pm$ & $0.97$ \\
	 $+$ &   $+$ &  cst. /  rot.  & $-0.67$ & $\pm$ & $0.40$ & $-0.63$ & $\pm$ & $0.63$ & $-0.53$ & $\pm$ & $0.98$ \\
	 $+$ &   $+$ &  rot. /  rot.  & $-0.71$ & $\pm$ & $0.43$ & $-0.65$ & $\pm$ & $0.62$ & $-0.7\phantom{0}$ & $\pm$ & $1.1\phantom{0}$ \\
	 $-$ &   $+$ &  rot. /  rot.  & $-0.79$ & $\pm$ & $0.44$ & $-0.77$ & $\pm$ & $0.68$ & $-0.68$ & $\pm$ & $0.97$ \\
	 $-$ &   $-$ &  rot. /  rot.  & $-0.86$ & $\pm$ & $0.41$ & $-0.78$ & $\pm$ & $0.67$ & $-0.73$ & $\pm$ & $0.87$ \\
	 $+$ &   $-$ &  rot. /  rot.  & $-0.88$ & $\pm$ & $0.47$ & $-0.79$ & $\pm$ & $0.68$ & $-0.9\phantom{0}$ & $\pm$ & $1.1\phantom{0}$ \\
	 $-$ &   $+$ &  cst. /  rot.  & $-1.01$ & $\pm$ & $0.44$ & $-1.03$ & $\pm$ & $0.62$ & $-1.1\phantom{0}$ & $\pm$ & $1.1\phantom{0}$ \\
	 $+$ &   $-$ &  cst. /  rot.  & $-1.22$ & $\pm$ & $0.45$ & $-1.10$ & $\pm$ & $0.68$ & $-1.2\phantom{0}$ & $\pm$ & $1.1\phantom{0}$ \\
	 $-$ &   $+$ &  rot. /  cst.  & $-1.23$ & $\pm$ & $0.46$ & $-1.11$ & $\pm$ & $0.72$ & $-1.09$ & $\pm$ & $0.98$ \\
	 $-$ &   $-$ &  cst. /  rot.  & $-1.71$ & $\pm$ & $0.43$ & $-1.65$ & $\pm$ & $0.65$ & $-1.71$ & $\pm$ & $0.96$ \\
	 $+$ &   $+$ &  rot. /  cst.  & $-1.97$ & $\pm$ & $0.48$ & $-1.88$ & $\pm$ & $0.69$ & $-1.87$ & $\pm$ & $0.99$ \\
	\hline
	\end{tabular} 
\end{table*}

Furthermore, the comparison with Jeffreys' scale (Table \ref{tab:jeffreys}) indicates that the $p(\gamma,K^{+})\Lambda$ data exhibit decisive evidence for the model variant with a positive vector and a negative tensor coupling constant, and a rotating phase for both trajectories. Indeed, the difference in $\ln\mathcal{Z}$ with the second-best model is around 24, amply exceeding the value of 5 required for a decisive statement. This result resolves the sign and phase ambiguity for $K^{+}\Lambda$ photoproduction, which previously could not be achieved using high-energy data alone~\cite{corthals-2006}.  
Moreover, the result from this Bayesian analysis is consistent with the previous analysis of this particular model, but it did not require an additional analysis with data from the resonance region (for which $E^{\gamma}_{lab} \lesssim 3$ GeV). In other words, the Nested Sampling method requires less experimental data to reach the same conclusion as the $\chi^2$-analysis which used a much larger set of $K^{+}\Lambda$ photoproduction data.

Apart from the relative probability of the different models, the Nested Sampling technique also provides us with an estimation value of the different parameters. In our best model, these values are:
\begin{align}
\frac{g_{K^+\Lambda p}}{\sqrt{4\pi}} &= -3.22 \pm 0.04 \nonumber\\
G_{K^{\ast +}}^{v} 	&= 12.47	\pm 0.14	\nonumber\\
G_{K^{\ast +}}^{t} 	&= -32.19	\pm 0.50.
\end{align}

Additional calculations demonstrate that the results do not change significantly when the $g_{K^+Y^{0}p}$ coupling constant is allowed to deviate up to 40\% from SU(3) predictions. For example, the values of $\Delta \ln{\mathcal{Z}}$ for the second and third $K^{+}\Lambda$ production models are $-24.21 \pm 0.73$ and $-77.09 \pm 0.70$ respectively, agreeing with the values found for 20\% SU(3) symmetry breaking. The estimation values for the coupling constants are not affected. We conclude that the high-energy $p(\gamma,K^{+})\Lambda$ data support a coupling constant $g_{K^+\Lambda p}$ compatible with a level of SU(3) symmetry breaking of at most 20\%. The estimation value for $g_{K^+\Lambda p}$ deviates 15\% from the SU(3) prediction.

\begin{figure}[]
 \centering
 \includegraphics[width=\columnwidth]{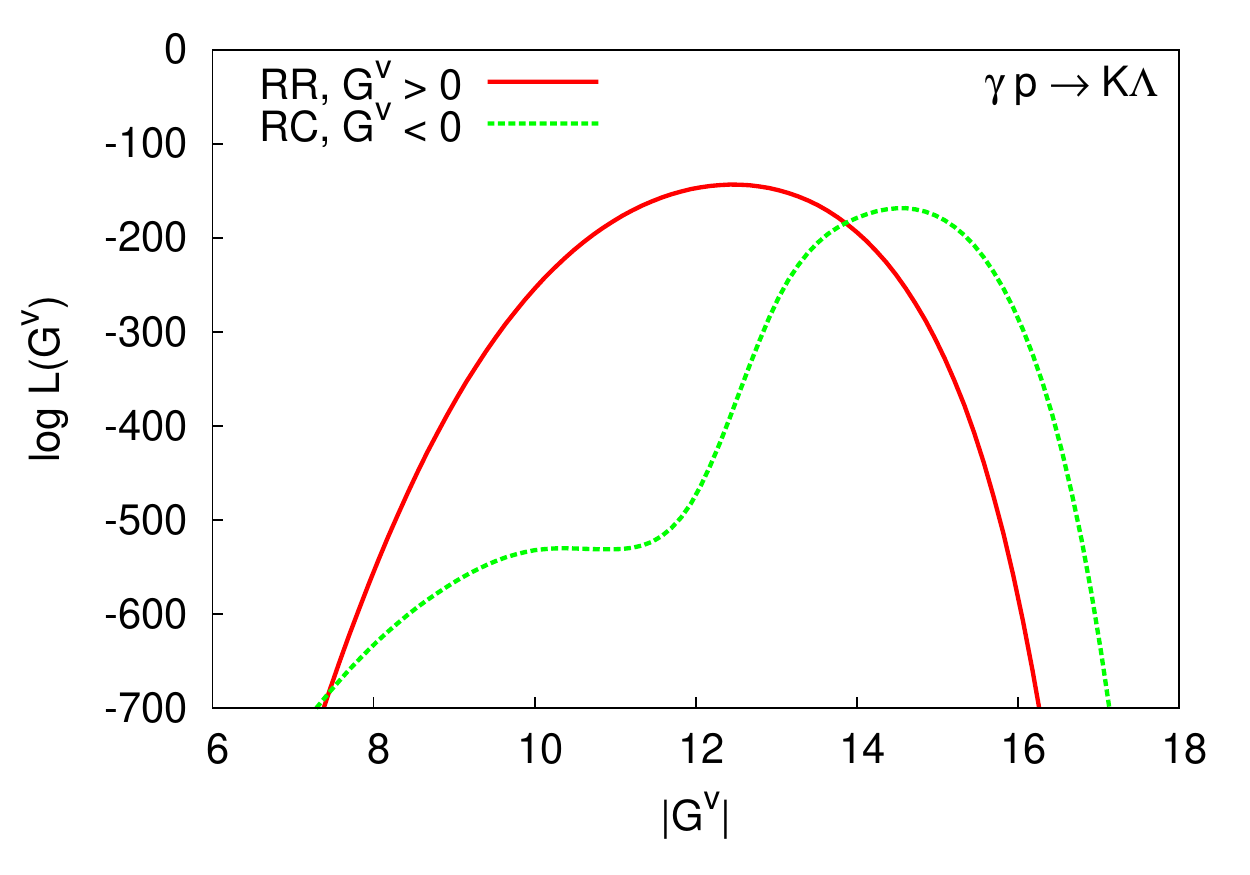}
 \caption{Comparison of the log-likelihood of $|G_{K^{\ast +}}^{v}|$ for the two best models. The model with the highest evidence is a rot./rot.~(RR) model with a positive estimation value for this parameter. The second-best model, rot./cst.~(RC), has a negative estimation value.}
 \label{fig:likelihood_KL}
\end{figure}

Fig. \ref{fig:likelihood_KL} shows the log-likelihood of the parameter $G_{K^{\ast +}}^{v}$, integrated over the remaining two parameters, using a uniform prior between -100 and 100. The likelihood is determined by the high-energy $K^+\Lambda$  photoproduction data only. The greater width and height of the peak of the rot./rot. model variant indicate that it should have a greater evidence than the alternative model.

The results for $K^+\Sigma^{0}$ production, listed in Table \ref{tab:lnz-ks}, are not as clear-cut as those for $p(\gamma,K^{+})\Lambda$. The difference in $\ln\mathcal{Z}$ of the model variants is next to negligible. This was to be expected, as a smaller dataset provides fewer restraints on the models' parameters. 

Even when the extensive set of resonance-region data is taken into account, the experimental data from the proton target alone does not allow us to single out one background model for the $K^+\Sigma^{0}$ channel~\cite{corthals-2007b}. However, a recent analysis of the corresponding reaction on the neutron, $n(\gamma,K^{+})\Sigma^{-}$, was able to resolve the remaining ambiguity~\cite{vancraeyveld-2009}. The models' parameters were converted from the proton to the neutron channel using isospin considerations.

\begin{figure}[]
 \centering
 \includegraphics[width=\columnwidth]{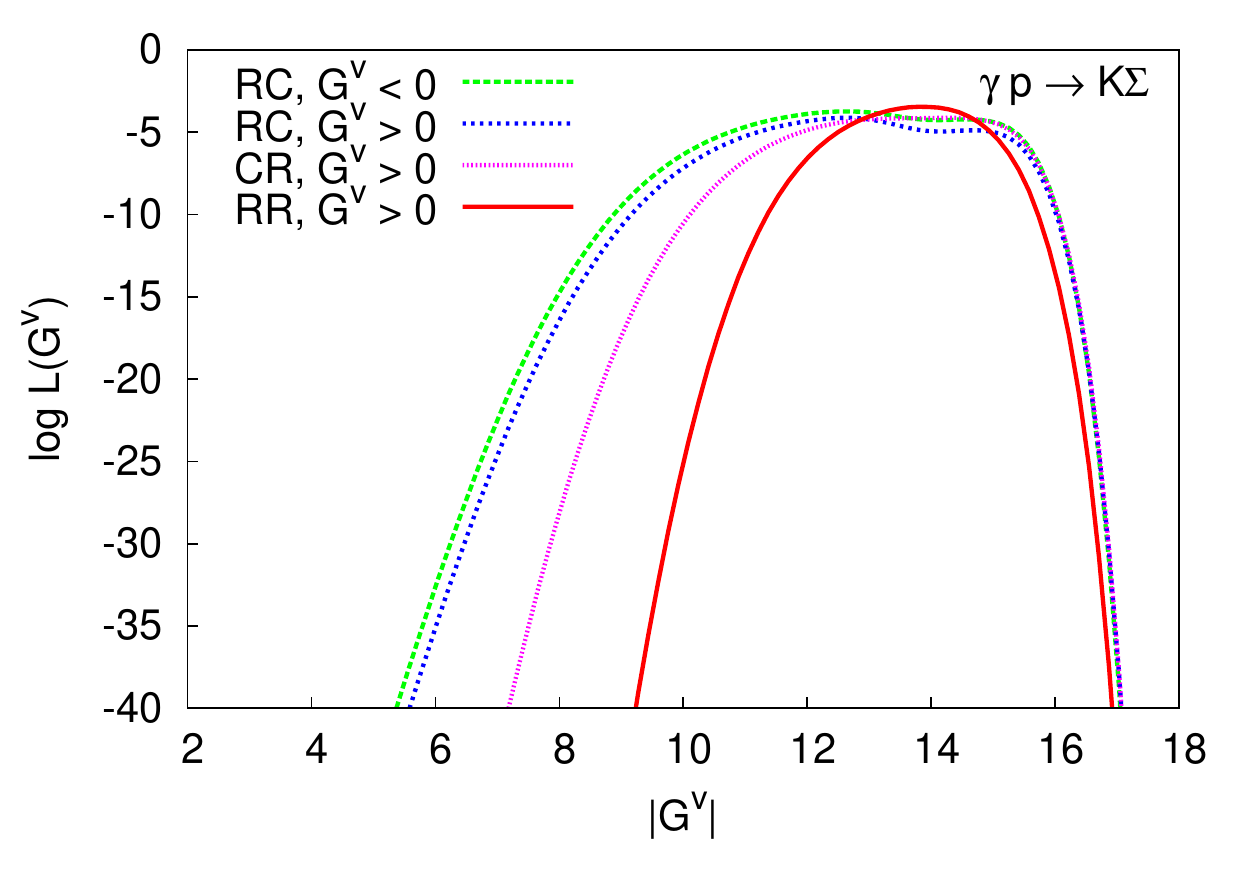} 
 \caption{Comparison of the log-likelihood of $|G_{K^{\ast +}}^{t}|$ for the four best model variants. In contrast with Fig. \ref{fig:likelihood_KL}, the model with the highest evidence value does not have the highest peak. Although the rot./cst.~(RC) models surpass the cst./rot.~(CR) and rot./rot.~(RR) variants, the difference between the evidences is still too small to make a decisive statement.
} 
 \label{fig:likelihood_KS}
\end{figure}

The log-likelihood of the parameter $G_{K^{\ast +}}^{v}$ for $K^+\Sigma^{0}$ production is shown in Fig. \ref{fig:likelihood_KS}. Note that despite boasting the maximal likelihood value, the rot./rot model has a lower evidence than the rot./cst. model. 

\section{Conclusions and outlook}\label{sec:conclusion}

Bayesian inference provides us with a promising tool for model comparison. We have demonstrated this by using the Nested Sampling algorithm to compute the Bayesian evidence for different model variants of a Regge model for $KY$ photoproduction. The results of this calculation indicate that there is decisive evidence for a $K^+\Lambda$ production model with $G^{v}_{K^{\ast+}} > 0$, $G^{t}_{K^{\ast+}} < 0$, and a rotating phase factor for both trajectories. This conclusion could not be drawn from the high-energy data set by means of the method of $\chi^2$ minimisation.

For $K^+\Sigma^{0}$ production, the differences in evidence are too small to draw a decisive conclusion, and supplementary data is required to fully determine the background model for this channel. 

The Nested Sampling method has many applications, both for the RPR model and for other research. One of these applications is the accurate estimation of model parameters as well as the elimination of nuisance parameters. More importantly, however, this method may provide us with a means to address the missing-resonance problem by calculating the probability of individual resonance contributions in a Bayesian framework. This is an approach we intend to explore in the near future.

\section*{Acknowledgements}

This research was funded by the Research Foundation - Flanders (FWO Vlaanderen). D.G.I. acknowledges the support of the UK Science and Technology Facilities Council.

\bibliographystyle{elsarticle-num}
\biboptions{sort&compress}
\bibliography{bayes-els}
\end{document}